\documentclass[11pt,twoside]{article}

\usepackage{asp2006}
\usepackage{fancyhdr,graphicx,caption,rotating,multirow,lscape}
\DeclareGraphicsExtensions{.eps,.eps.gz,.ps,.jpg,.tiff}

\begin{document}

\title{Signal Search and Reconstruction by a Trend Filtering Algorithm}  

\author{G.~Kov\'acs}  
\affil{Konkoly Observatory, Budapest, Hungary}
\author{G.~\'A.~Bakos}
\affil{Harvard-Smithsonian Center for Astrophysics, Cambridge, USA}

\begin{abstract}
We present additional tests of our algorithm aimed at filtering out 
systematics due to data reduction and instrumental imperfections in 
time series obtained by ensemble photometry. Signal detection 
efficiency is demonstrated, and a method of decreasing the false 
alarm probability is presented. Including the recently discovered 
transiting extrasolar planet HAT-P-1, we show various examples on the signal 
reconstruction capability of the method. 
\end{abstract}


\section{Introduction}
One of the most serious challenges in the hunt for transiting 
extrasolar planets is the removal of the various systematics 
remaining in the photometric databases even after employing 
sophisticated methods of CCD image reduction. Sometimes called 
as ``red noise'' \citep{pzq2006}, systematics/trends may show 
up in many ways. Their most common appearance is a drift with 
$\sim 1$~d$^{-1}$ frequency due to variations of the point-spread 
function and numerous other parameters, for example focus 
change, sub-pixel coordinate drifts, etc. In addition to these 
most trivial systematics, we may have many others, from transients 
(e.g., imperfect removal of cosmic rays) 
to periodic saturation of bright stars, depending on Moon phase. 
During the past several years it has been realized that, except 
for the simple, high signal-to-noise ratio cases, filtering out 
these systematics from the databases is absolutely crucial from 
the point of view of transit search. Therefore, efforts have been 
taken to devise post processing algorithms that are capable of 
whitening out the data from the systematics. There are two methods 
in this field that attracted wider interest\footnote{A third method 
by \citet{ks2003}, that employs straightforward iterative Fourier 
filtering, has been used apparently less frequently.}. The method 
SysRem by \citet{tmz2005} uses basically an iterative principal 
component analysis to filter out the most prominent systematics 
from the data. The Trend Filtering Algorithm (TFA) by \citet{kbn2005} 
is a least-squares method that is capable of filtering out nearly 
arbitrary systematics, assuming that the selected set of templates 
is ``flavorous'' enough, i.e., it contains the light curves necessary 
for the approximation of the type of trend observed in the target. 
Here we present tests on the signal recovery capability of TFA. 
%
%
%
\section{The Algorithm}
For completeness, we briefly summarize the basic steps of TFA. First 
we select a {\em template set} of $M$ light curves from the photometric 
database of the field of interest. From these 
\{$X_j(i); j=1,...,M; i=1,...,N$\} time series (sampled in $N$ moments 
of time), we construct the following filter: 
%
%
\begin{eqnarray}
F(i) & = & \sum_{j=1}^M c_j X_j(i) \hskip 2mm .
\end{eqnarray}
The coefficients \{$c_j$\} for a target \{$Y(i)$\} are determined 
by minimizing the following expression: 
%
%
\begin{eqnarray}
\cal D & = & 
\sum_{i=1}^N [Y(i) - A(i) - F(i)]^2 \hskip 2mm .
\end{eqnarray}
Here the function $A(i)$ is derived in the following way:  
%
%
\[ A(i) = \left\{
\begin{array}{ll}
\langle Y\rangle = const  & 
\hskip 8mm\mbox{; for period search} \hskip 2mm , \\
A(i) \Leftrightarrow Y(i) - F(i) &
\hskip 8mm\mbox{; for signal reconstruction} \hskip 2mm .
\end{array}   
\right. \makebox[23mm][r]{(3)}  \] 
Namely, in the case of period search we assume that the observed 
signal is dominated by systematics and noise, and therefore, the 
filter is expected to yield minimum dispersion around the constant 
signal average $\langle Y\rangle$\footnote{If the observed signal 
is dominated by the true signal, the above approximation for 
frequency search is still applicable, since the temporal behavior 
of the systematics is still unlikely to be the same as that of the 
signal. In both the systematics- and signal-dominated cases the 
true signal suffers from some distortion, but in the latter case 
the distortion will become more obvious.
}. Once the signal is identified, we can recover its shape by 
iteratively approximating the noiseless and trend-free signal 
\{$A(i)$\}. (The iteration is indicated by the symbol 
$\Leftrightarrow$ in Eq.~(3).) In this iterative process we assume 
that \{$A(i)$\} can be represented by a low-parameter model, and 
that the observed signal minus the systematics yield the true signal 
with white noise. 
%
%
%
\section{Transit Detection Efficiency}
The datasets used in this paper are listed in Table~1. We test 
transit detection efficiency (TDE), false alarm probability (FAP) 
and signal reconstruction capability (SRC). For testing SRC  
on datasets different from those of HATNet \citep{bbbb2004}, we 
use the HAT-P-1 \citep{bbbb2006} observations by the 60/90/180cm 
Schmidt telescope of the Konkoly Observatory.  

To show that TFA is capable of filtering out nearly all systematics, 
we compute the distribution function of the peak frequencies obtained 
by the BLS analysis \citep{kzm2002}. The result of this analysis for 
field \#125 is shown in Fig.~1. We see that the original data includes 
many stars with various periodic systematics, some of which are not 
too easy to relate to the daily change of the observational conditions. 
For example, the strongest peak at $\sim 0.16$~d$^{-1}$ is most 
probably associated with the saturation of the bright stars, because 
after omitting the first $\sim 400$ stars, the peak at this frequency 
becomes less prominent than the ones corresponding to other systematics. 
%
%
\begin{table}[h]
\smallskip
\caption{Properties of the test datasets}
\begin{center}
{\small
\begin{tabular}{lrrrrl}
\tableline
\noalign{\smallskip}
Set & $N$ & $N_{\rm star}$ & $T$[d] & I [mag] & Purpose \\
\noalign{\smallskip}
\tableline
\noalign{\smallskip}
HATnet \#125     & $6100$ & $24980$ & $141.0$ & $6.3-9.9$  & TDE, FAP \\
HATnet \#127     & $2100$ & $13620$ & $167.0$ & $7.6-9.7$  & FAP      \\
HATnet \#148     & $5000$ & $ 3430$ & $112.0$ & $7.4-13.6$ & SRC      \\
HAT-P-1, Schmidt & $650 $ & $  202$ & $0.32 $ & $9.6-16.5$ & SRC      \\ 
\noalign{\smallskip}
\tableline
\end{tabular}
}
\end{center}
\footnotesize
\underline{Notes:}\hspace{2mm}
\parbox[t]{120mm}
{$N$: number of data points per object;
$N_{\rm star}$: number of stars in the field;
$T$: total time span; 
I [mag]: I magnitude range in the field. 
}
\end{table}
%
%
%
\begin{figure}[h]
\centering
\begin{minipage}[c]{.49\textwidth}
\includegraphics[width=60mm]{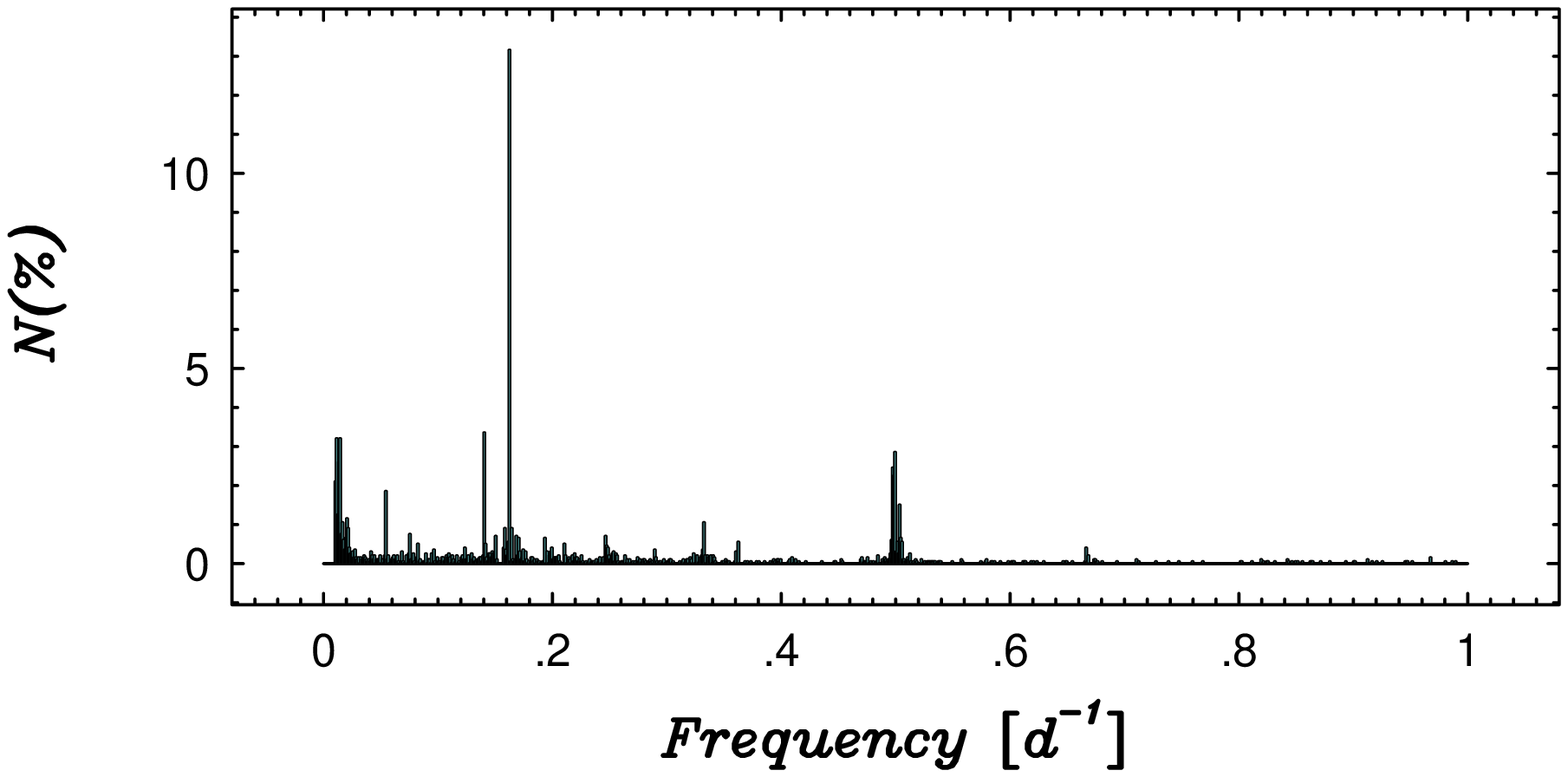}
\end{minipage}
\begin{minipage}[c]{.49\textwidth}
\includegraphics[width=60mm]{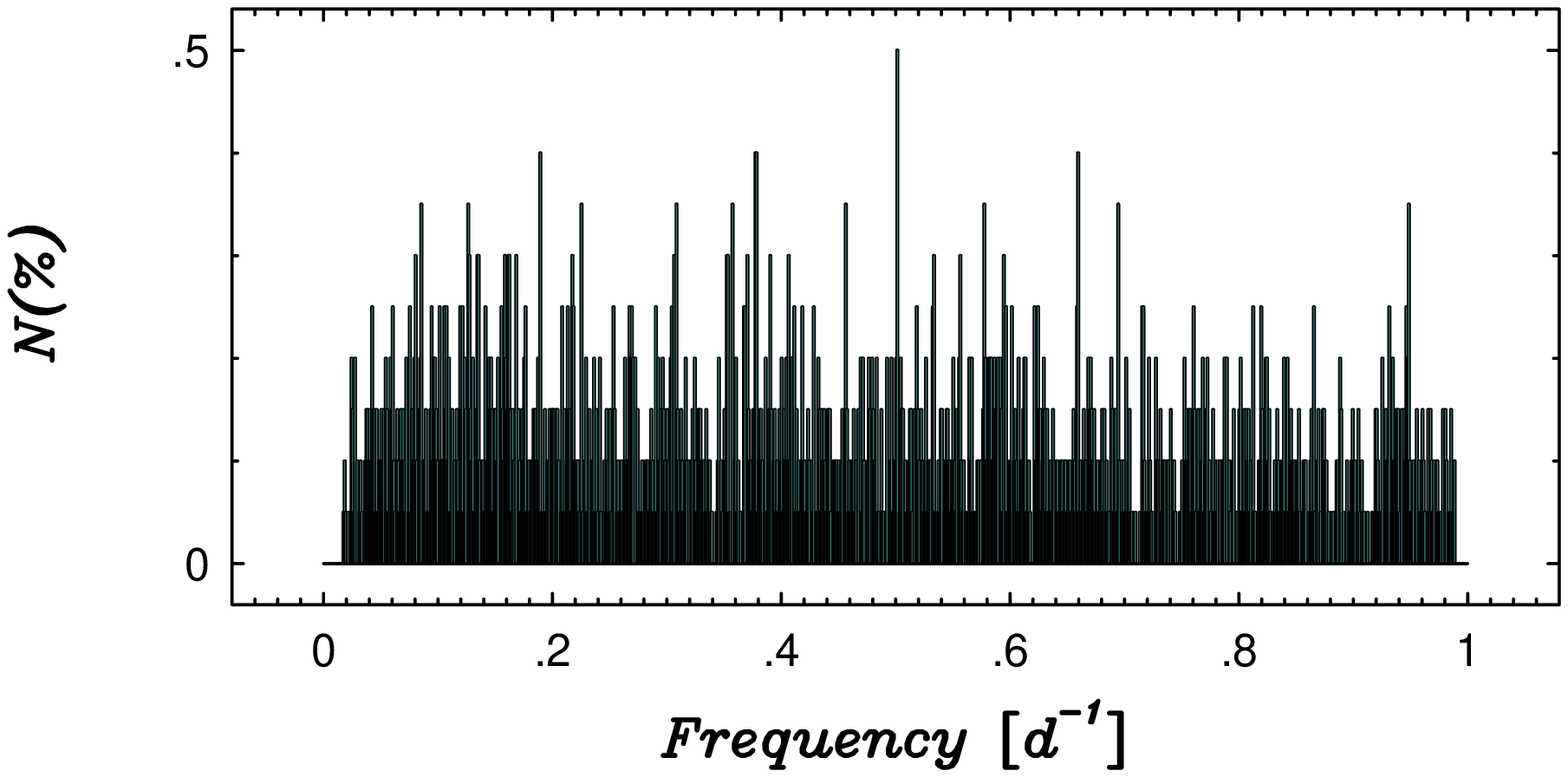}
\end{minipage}
\caption{ : Distribution of the peak frequencies of the 
            BLS spectra of the first 2000 brightest stars in 
	    HATNet field \#125. Left panel: raw (original) data; 
	    right panel: TFAd data with 900 templates. Please 
	    note that the application of TFA leads to a nearly 
	    flat distribution of the peak frequencies as expected 
	    for white noise signals.
\label{fredi13}}
\end{figure}

We have already demonstrated in Kov\'acs et al. (2005) the ability 
of TFA of detecting shallow transits that are buried in noise and 
systematics. Because the statistics we use have slightly changed 
from then, here we show the results of tests conducted with the new 
statistics. 

We inject a periodic transit signal in the given target from the 
first $2000$ stars of field \#125. Then we run a TFA/BLS analysis on 
the target and check if the DSP parameter \citep{kb2005} corresponding 
to the highest peak in the frequency spectrum exceeds a given limit. 
The DSP parameter expresses the significance of the dip corresponding 
to the transit derived from the analysis. In order to exclude binaries with 
light reflection and gravitational effects, DSP also includes 
weighting by the most significant Fourier component of the 
out-of-transit variation. When computing DSP, we always use the 
TFA code in the signal reconstruction mode to get a better estimate 
on this parameter. 

Properties of the injected signal and the number of detections are 
listed in Table~2. We note that the synthetic signal has a flat 
out-of-transit part and a trapeze shape with rather long 
ingress/egress durations. The condition of detection is given by a 
cutoff imposed on DSP. The value of this cutoff is large enough to 
eliminate false alarms (see Sect.~4). We see that there is a highly 
significant increase in the detection probability due to the 
application of TFA. This increase is especially striking when the 
top 500 bright stars are tested. These are the ones that are most 
seriously affected by systematics related to saturation effects 
(see above). We note that the detection ratio can be slightly increased 
(from $46$\% to $50$\%) if we choose templates only from this brighter 
set of stars.    
%
%
\begin{table}[h]
\smallskip
\caption{Detection of injected transit signals in field \#125}
\begin{center}
{\small
\begin{tabular}{crrc}
\tableline
\noalign{\smallskip}
TFA & $N_{\rm star}$ & $N_{\rm d}$ & $N_{\rm d}$[\%] \\
\noalign{\smallskip}
\tableline
\noalign{\smallskip}
0  & $2000$ & $ 972$  & $49$ \\
1  & $2000$ & $1340$  & $67$ \\
0  & $ 500$ & $  51$  & $10$ \\
1  & $ 500$ & $ 228$  & $46$ \\
\noalign{\smallskip}
\tableline
\end{tabular}
}
\end{center}
\footnotesize
\underline{Notes:}\hspace{2mm}
\parbox[t]{120mm}
{Analysis: BLS, with minimum transit duration of $0.01P_{\rm {test}}$, 
$P_{\rm {test}}\in[1.0,100.0]$~d; 
Detection condition: DSP$>8.0$;  
Injected signal parameters: period, $P=5.123$~d;  
fractional transit length, $q_{\rm tran}=\Delta t_{\rm tran}/P=0.02$; 
fractional ingress length, 
$q_{\rm ingr}=\Delta t_{\rm ingr}/\Delta t_{\rm tran}=0.40$;  
transit depth, $d=-0.015$~mag.
}
\end{table}
%

%
%
\section{False Alarms}
By ``false alarms'' we mean those cases when the detection statistics 
indicate the presence of a signal, but the probability distribution 
of the statistics (derived on pure noise) shows that the observed 
value may also occur due to a random event and the probability of 
this to happen is ``greater than we would like to''. In assessing FAP 
we resort to direct statistical tests, in which we generate pure 
Gaussian time series on the time base of the observed light curves. 
We analyze these artificial time series and count the number of cases 
when DSP exceeds a prescribed limit. In this way we get estimates on 
FAPs for a given dataset when TFA and BLS are used. 

The results of the tests are presented in Table~3. Several 
conclusions can be drawn from this table. First of all, 
as expected, application of TFA introduces correlation 
in the time series.\footnote{
This ``side effect'' is unavoidable in any data fitting. 
In the applications the net result will depend on the relative 
weight of this correlation to the one introduced by the systematics.} 
This increases FAP by a substantial amount. Second, larger number 
of data points results in a decrease of FAP. Third, larger number 
of templates leads to stronger correlation, and therefore, to 
an increase of FAP. Fourth, although increasing the number 
of data points decreases FAP, it is present in a relative 
high value even for higher DSP cutoff values (when we already 
expect a visible sign of the (fake) transit in the 
folded/binned light curve\footnote{It is noted, however, 
that in many cases the high/moderate DSP detections are 
due to short events containing few data points.}. 

In an attempt to reduce FAP, the following method is suggested. 
By using several TFA runs, corresponding to various template 
numbers, we compute DSP values for the given database. Due to 
the way the template sets are constructed, these results are 
expected to be largely independent from each other. In a 
conservative approach of signal detection, we require the signal 
to be present in all these runs. In the primary selection of 
transit candidates we require only that the dip is negative 
(i.e., corresponding to dimming) and that ${\rm DSP}>{\rm DSP}_{\rm cut}$. 
The result of this multiple template FAP filtering is shown 
in the lower three lines of Table~3. We see that the method is 
very effective already with three different TFA runs. For example, 
even for the sparsely sampled field \#127, with three or more 
different TFA runs we can filter out false alarms with a 
probability better than 99.9\% for signals with DSP$>7$. 
%
%
\begin{table}[h]
\smallskip
\caption{Testing false alarms in fields \#125 and \#127}
\begin{center}
{\small
\begin{tabular}{lrrrcrrr}
\tableline
\noalign{\smallskip}
\multicolumn{1}{c}{TFA} & 
\multicolumn{3}{c}{$N_{\rm d}(\#127)$} & 
\multicolumn{1}{c}{} & 
\multicolumn{3}{c}{$N_{\rm d}(\#125)$} \\
\noalign{\smallskip}
\tableline
${\rm DSP}_{\rm cut}$\ : & 5 & 6 & 7 & & 5 & 6 & 7 \\
\noalign{\smallskip}
\tableline
\noalign{\smallskip}
$\emptyset$ &  68 &   7 &   0 & & 26  &  3 & 1 \\
a           & 827 & 400 & 107 & & 258 & 24 & 2 \\
b           & 856 & 523 & 168 & & 306 & 31 & 6 \\
c           & 842 & 560 & 220 & & 394 & 42 & 3 \\
d           & 890 & 638 & 328 & & 474 & 60 & 5 \\
\noalign{\smallskip}
\tableline
ab          & 386 & 112 &   9 & &  73 &  2 & 0 \\
abc         & 184 &  38 &   1 & &  25 &  0 & 0 \\
abcd        &  97 &  21 &   1 & &  14 &  0 & 0 \\
\noalign{\smallskip}
\tableline
\end{tabular}
}
\end{center}
\footnotesize
\underline{Notes:}\hspace{2mm}
\parbox[t]{120mm}{
Test signal: pure Gaussian noise; datasets: the top $2000$ bright 
stars in each field;  
Analysis: BLS, with minimum transit duration of 
$0.02P_{\rm {test}}$, $P_{\rm {test}}\in[1.0,100.0]$~d;  
TFA template numbers: 0, 700, 800, 900 and 1000 for $\emptyset$, 
a, b, c and d, respectively; 
DSP lower detection limits: 5, 6 and 7;  
Items in the table show the number of detections 
(negative dips with ${\rm DSP}\geq{\rm DSP}_{\rm cut}$); 
ab, abc, abcd: datasets used for finding simultaneous detections.
}
\end{table}
%

%
%
\section{Signal reconstruction}
Signal reconstruction is an essential (but optional) part of signal processing 
when TFA is used. This is because we do not know {\em a priori} which 
part of the observed signal comes from the systematics and which 
one from the true signal (and all these are coupled with noise). 
Without knowing the signal parameters {\em a priori}, we resort to an 
iterative scheme in reconstructing the true signal (see also 
Sect.~2). Our experiments on the HATNet database show that 
this reconstruction can be quite successful without making any 
assumption on the signal shape. Once the signal shape is reliably 
identified, one can proceed by more specific assumptions, e.g., 
by using trapeze transit shapes, and thereby further decreasing 
the number of parameters fitted. To illustrate the efficiency of 
the TFA reconstruction, we show two examples in Fig.~2.   
%
%
\begin{figure}[h]
\centering
\begin{minipage}[c]{.49\textwidth}
\centering
\includegraphics[width=65mm]{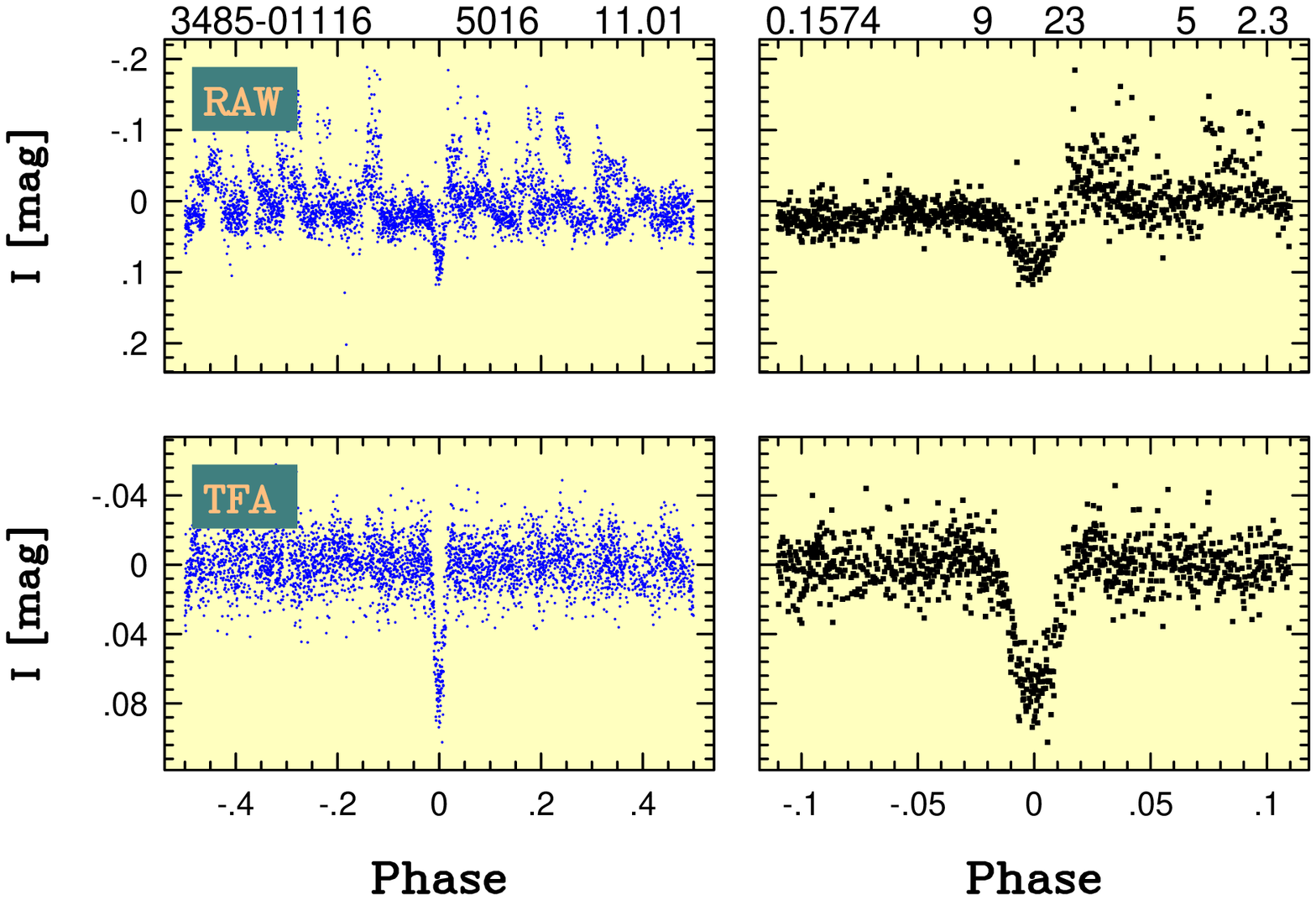}
\end{minipage}
\begin{minipage}[c]{.49\textwidth}
\centering
\includegraphics[width=55mm]{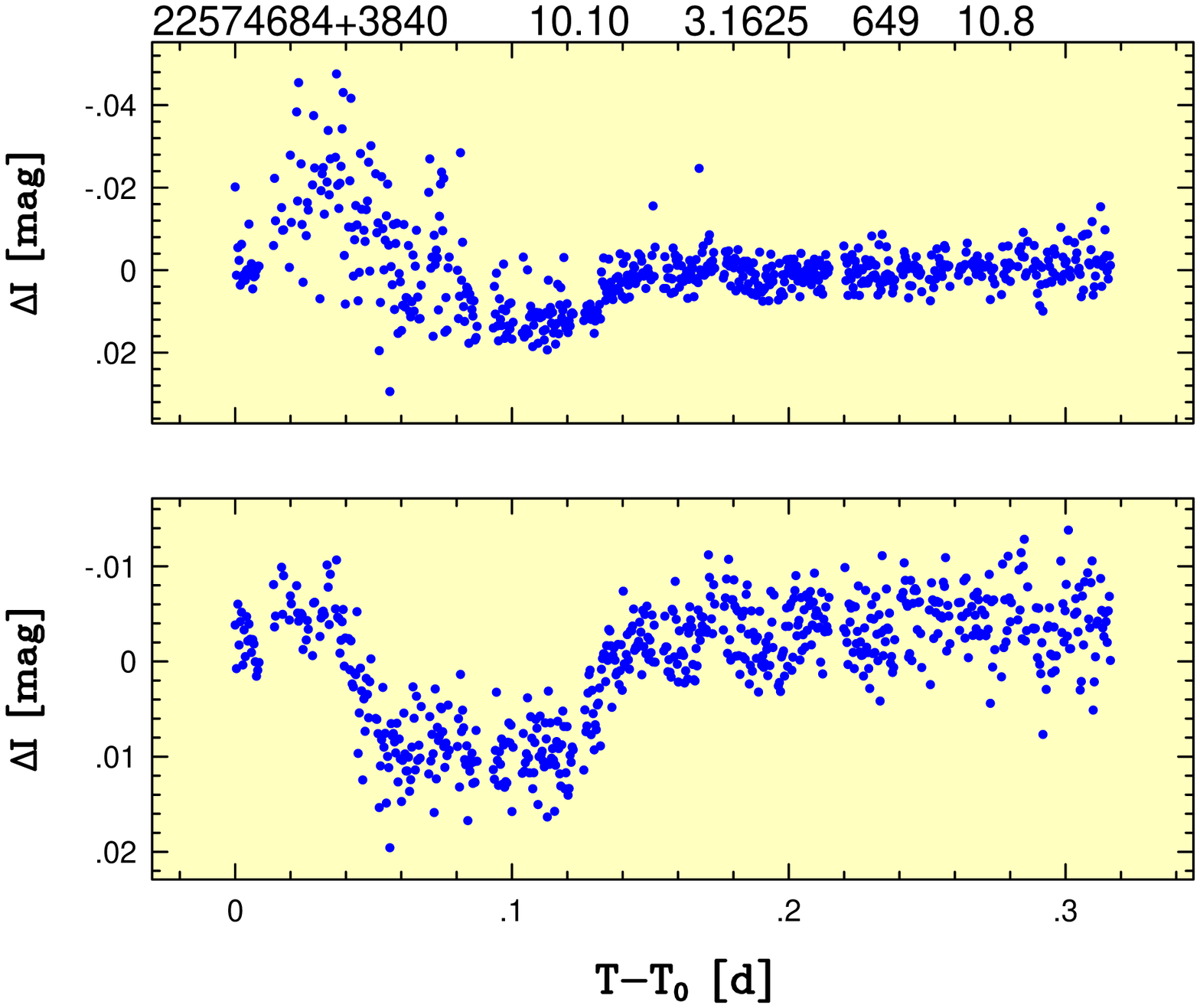}
\end{minipage}
\caption{ : Upper panels: ensemble photometry of an eclipsing variable 
            in field \#148 (left) and that of HAT-P-1 (right); 
	    lower panels: TFA-reconstructed light curves of the same 
	    objects. Headers (from left to right): star ID, number of 
	    data points, average I magnitude, main BLS frequency 
	    [d$^{-1}$], signal-to-noise ratio (SNR) of the BLS spectrum, 
	    DSP, SNR of the out-of-transit variation and its peak 
	    frequency in the units of the BLS frequency. On the right 
	    we have: star ID, average magnitude, plotting frequency 
	    [d$^{-1}$], number of data points, DSP. The reconstruction 
	    of HAT-P-1 was made without using its nearby companion star 
	    {\mbox ADS16402 A}. In both cases no assumptions were made 
	    on the signal shape. 
	    \label{tfarec}}
\end{figure}
%

%
%
\section{Conclusions}
Filtering out systematics from astrophysical time series 
is nearly mandatory if a survey-type analysis is made with 
the goal of reaching the theoretical white noise limit of 
signal detection. In the search for extrasolar transiting 
planets this issue becomes even more highlighted due to the 
delicacy of the detection. We have shown in this paper that 
TFA is capable of filtering out various systematics, thereby 
allowing the detection and a concomitant reconstruction of 
faint regular (e.g., simple- or multi-periodic) signals. 
Furthermore, by requiring multiple detections in time series 
filtered by various TFA templates, false alarm probability 
can be pushed down near to the white noise limit.  
  
\vspace{-2mm}

\acknowledgements 
Support for program number HST-HF-01170.01-A to G.~\'A.~B.~was provided by
NASA through a Hubble Fellowship grant from the Space Telescope Science
Institute. Operation of the HATNet project is funded in part by NASA grant
NNG04GN74G. We also acknowledge OTKA K-60750.


\begin{thebibliography}{}
\bibitem[Bakos et al.(2004)]{bbbb2004}
Bakos, G.~\'A., Noyes, R.~W., Kov\'acs, G., et al.~2004, PASP, 116, 266
\bibitem[Bakos et al.~(2006)]{bbbb2006}
Bakos, G.~\'A., Noyes, R.~W., Kov\'acs, G., et al., 
2006, ApJ, (in press), (astro-ph/0609369)
\bibitem[Kov\'acs \& Bakos (2005)]{kb2005}
Kov\'acs, G.~\& Bakos, G.~\'A.~2005, poster paper (astro-ph/0508081) 
\bibitem[Kov\'acs, Bakos, \& Noyes (2005)]{kbn2005}
Kov\'acs, G., Bakos, G.~\'A., \& Noyes, R.~W.~2005, MNRAS, 356, 557  
\bibitem[Kov\'acs, Zucker \& Mazeh (2002)]{kzm2002}
Kov\'acs, G., Zucker, S.~\& Mazeh, T.~2002, A\&A, 391, 369
\bibitem[Kruszewski \& Semeniuk (2003)]{ks2003}
Kruszewski A., Semeniuk I.~2003, Acta Astr., 53, 241
\bibitem[Pont, Zucker \& Queloz (2006)]{pzq2006}
Pont, F., Zucker, S., \& Queloz, D.~2006, MNRAS, 373, 231  
\bibitem[Tamuz, Mazeh \& Zucker (2005)]{tmz2005}
Tamuz, O., Mazeh, T., \& Zucker 2005, MNRAS, 356, 1466
\end{thebibliography}
\end{document}